# Chapter 5

# Collimation System

*R.B. Appleby[1], R. Barlow[2], A. Bertarelli[3], R. Bruce[3], F. Carra[3], F. Cerutti[3], L. Esposito[3], A. Faus-Golfe[4], H. Garcia Morales[5], L. Gentini[3], S. M. Gibson[5], P. Gradassi[3], J.M. Jowett[3], R. Kwee-Hinzmann[5], L. Lari[6], A. Lechner[3], T. Markiewicz[7], A. Marsili[3], J. Molson[1], L.J. Nevay[5], E. Quaranta[3], H. Rafique[2], S. Redaelli[3]\*, M. Serluca[1], E. Skordis[3], G. Stancari[8], G. Steele[3] and A. Toader[2]*

[1]UMAN, The University of Manchester and the Cockcroft Institute, Warrington, UK
[2]UHUD, University of Huddersfield, Huddersfield, UK
[3]CERN, Accelerator & Technology Sector, Geneva, Switzerland
[4]IFIC, Instituto de Fisica Corpuscular, Valencia, Spain
[5]URHL, Royal Holloway, London, UK
[6]IFIC, Instituto de Fısica Corpuscular, Valencia, Spain (now ESS European Spallation Source, Lund, Sweden
[7]SLAC National Accelerator Laboratory, Menlo Park, USA
[8]FNAL, Fermi National Accelerator Laboratory, Batavia, USA

## 5 Collimation system

### 5.1 LHC multi-stage collimation system

#### 5.1.1 Motivation

A variety of processes can cause unavoidable beam losses during normal and abnormal operation. Because of the high stored energy above 700 MJ, the beams are highly destructive. Even a local beam loss of a tiny fraction of the full beam in a superconducting magnet could cause a quench, and large beam losses could cause damage to accelerator components. Therefore, all beam losses must be tightly controlled. For this purpose, a multistage collimation system has been installed [1–8] to safely dispose of beam losses. Unlike other high-energy colliders, where the main purpose of collimation is to reduce experimental background, the LHC and the HL-LHC require collimation during all stages of operation to protect its elements.

The HL-LHC imposes increased challenges for the collimation system. For the same collimation cleaning and primary beam loss conditions as in the LHC, the factor ~2 increase in total stored beam energy foreseen from the HL-LHC parameters requires a corresponding improvement of cleaning performance to achieve the same losses in cold magnets. Total losses might also exceed the robustness limit of collimators. The LHC system was designed to safely withstand beam lifetime drops down to 0.2 h during 10 s, corresponding to peak losses of up to 500 kW. As mentioned above, these loss levels scale with the total beam intensity so they will increase by about a factor of 2 for the HL-LHC parameter set. The collimation system must be upgraded to cope with these higher loss levels. It is clear that the lifetime control and optimization of beam halo losses will be crucial for the LHC upgrade (see also Section 5.4 Other collimators of the present system required in the HL-LHC). The larger stored energy, together with smaller beam sizes achieved through higher brightness beams, also imposes more severe challenges for collimator robustness against design loss scenarios for cleaning. In the case of single-turn beam failures, brighter beams significantly increase thermo-mechanical loads on collimator materials and components. The higher peak luminosity challenges entail the definition of new concepts for physics debris cleaning and an overall redesign of the IR collimation layouts. For example, in the present LHC layout, the inner triplet represents the IR aperture bottleneck and is protected by two dedicated tertiary collimators per plane per beam. Future optics scenarios might add critical aperture restrictions at magnets further away from the IP, requiring additional cleaning and protection.

---

[*] Corresponding author: Stefano.Redaelli@cern.ch

To meet the new challenges, the HL-LHC collimation system therefore builds on the existing LHC collimation system, with the addition of several upgrades.

### 5.1.2   Collimation system inherited from the LHC

The backbone of the HL-LHC collimation system will remain, as for the current LHC, the betatron (IR7) and momentum (IR3) cleaning systems installed in two separated warm insertions [1]. A very efficient halo cleaning, as required to operate the LHC with unprecedented stored beam energies in a superconducting collider, is achieved by very precisely placing blocks of materials close to the circulating beams, while respecting a pre-defined collimator hierarchy that ensures optimum cleaning in a multi-stage collimation process. This is illustrated schematically in Figure 5-1. Most collimators consist of two movable blocks referred to as 'jaws', typically placed symmetrically around the circulating beams. The present system deployed for LHC operation between 2010 and 2013 provided a cleaning efficiency of above 99.99% [2], i.e. it ensured that less than $10^{-4}$ of the beam losses are lost in superconducting magnets.

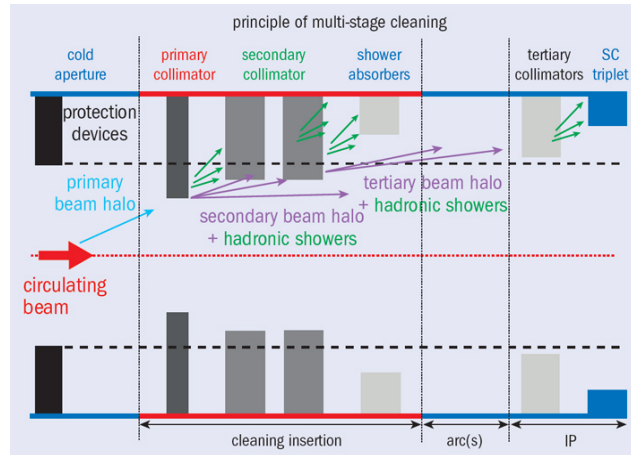

Figure 5-1: Schematic illustration of multi-stage collimation cleaning at the LHC. Primary and secondary collimators (darkest grey) are the devices closest to the circulating beam and are made of robust carbon-fibre composites. Shower absorbers and tertiary collimators (lighter grey) sit at larger apertures and are made of a tungsten alloy to improve absorption. Collimators of different families are ordered in a pre-defined collimation hierarchy that must be respected in order to ensure the required system functionalities. The collimator hierarchy is ensured by defining collimator settings in units of local beam size at the collimator location.

The LHC collimators are built as high-precision devices with beam sizes as small as 200 microns, in order to ensure the correct hierarchy of devices along the 27 km ring. Details of the collimator design can be found in Ref. [9]. Key features of the design are (i) a jaw flatness of about 40 microns along the 1 m long active jaw surface; (ii) a surface roughness below 2 microns; (iii) a 5 micron positioning resolution (mechanical, controls); (iv) an overall setting reproducibility below 20 microns [10]; (v) a minimal gap of 0.5 mm; and (vi) evacuated heat loads of up to 7 kW in a steady-state regime (1 h beam lifetime) and of up to 30 kW in transient conditions (0.2 h beam lifetime). Two photographs of the present LHC collimator are given in Figure 5-2, where a horizontal collimator and a 45° tilted collimator are shown. An example of the tunnel installation layout for an IR7 collimator is given in Figure 5-3. The complete list of collimators, including injection protection collimators in the transfer lines, is given in Table 5-1. For completeness, the injection protection TDI blocks and the one-side beam dump collimator TCDQ (considered to be part of beam transfer rather than of the LHC collimation system, but designed with a similar concept) are also listed (see Chapter 14). The full system comprises 118 collimators, 108 of which are movable.

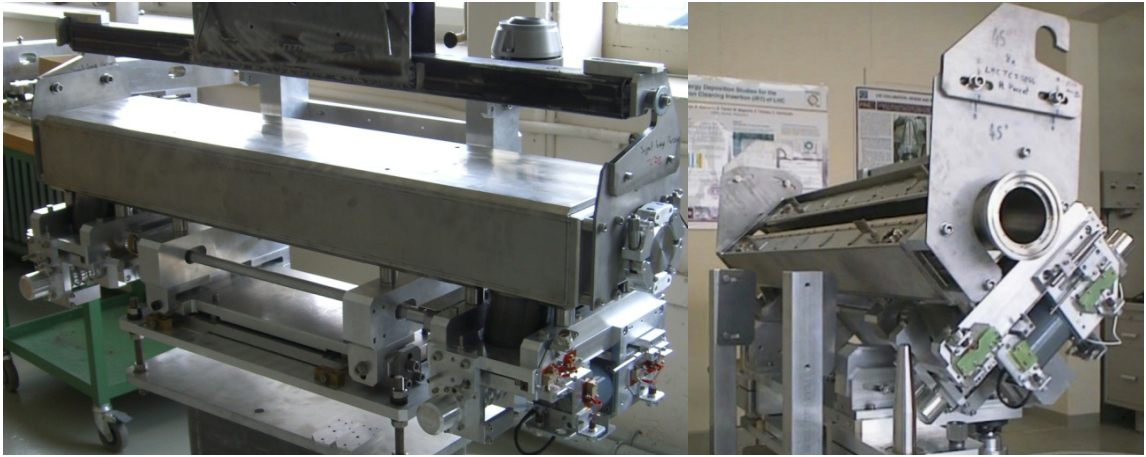

Figure 5-2: (a) horizontal LHC collimator; (b) skew LHC collimator. The latter has the vacuum tank open to show the two movable CFC jaws.

Since the collimator jaws are close to the beam (e.g. the minimum collimator gap in 2012 was 2.1 mm, i.e. jaws were 1.05 mm from the circulating beam), the collimation system also has a critical role in the passive machine protection in case of beam failures that cannot be counteracted by active systems. Primary and secondary collimators in IR7 are the closest to the beam; their jaws are mainly made of robust carbon-fibre carbon composites (CFC), and are designed to withstand beam impacts without significant permanent damage from the worst failure cases such as impacts of a full injection batch of $288 \times 1.15 \times 10^{11}$ protons at 450 GeV and of up to $8 \times 1.15 \times 10^{11}$ protons at 7 TeV [11]. However, they contribute significantly to the machine impedance, which is particularly critical at top energy, because of the low electrical conductivity of the CFC. Impedance constraints determine the smallest gaps that can be used in IR3 and IR7 and hence the minimum $\beta^*$ in the experiment [3]. Other absorbers and tertiary collimators are positioned at larger gaps in units of the local beam size. They can be less robust, in term of resistance to impact of the beam, than primary and secondary collimators because they are less exposed to beam losses. Thus, metal-based jaws that are more effective in absorbing particles can be used.

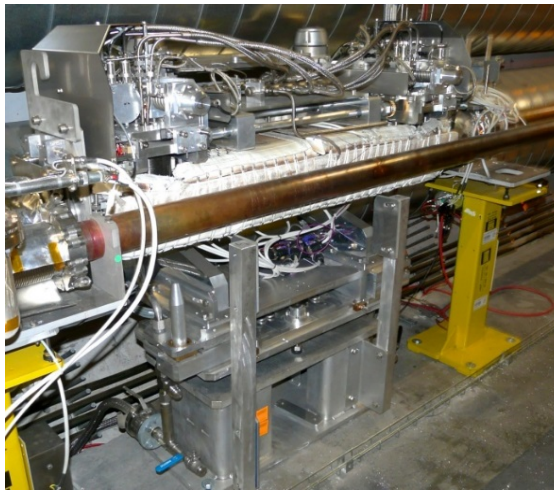

Figure 5-3: Photograph of the active absorber TCLA.B6R7.B1 as installed in the betatron cleaning insertion.

The initial collimator design has been improved by adding two beam position monitors (BPMs, known as 'buttons') on both extremities of each jaw [12]. Eighteen collimators (16 TCTP and 2 TCSP) have been already upgraded with this new design during LS1. This concept allows for fast collimator alignment as well as a continuous monitoring of the beam orbit at the collimator, while the BLM-based alignment can only be performed during dedicated low-intensity commissioning fills. The BPM buttons will improve significantly the collimation performance in terms of operational flexibility and $\beta^*$ reach [3]. The BPM collimator design is considered to be the baseline for future collimation upgrades, and the BPM design is equally applicable to all collimators regardless of the jaw material. The concept has been tested extensively at the CERN SPS with a collimator prototype with BPMs [13–15]. An example of a CFC jaw prototype with in-jaw BPMs is shown in Figure 5-4.

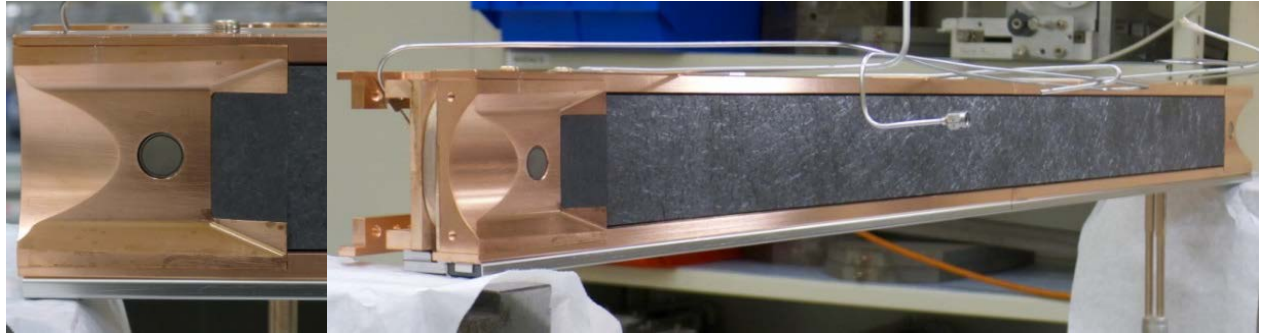

Figure 5-4: New carbon/carbon collimator jaw with integrated BPMs at each extremity ('buttons') to be installed as the secondary collimator in the dump insertion IR6. A detail of the BPM is given on the left hand side. A variant of this design, made with a Glidcop support and tungsten inserts on the active jaw part, will be used for the tertiary collimators in all IRs.

In addition to beam halo cleaning, the collimation system has other important roles.

- Passive machine protection: the collimators are the closest elements to the circulating beam and represent the first line of defense in case of various normal and abnormal loss cases. Due to the damage potential of the LHC beams, this functionality has become one of the most critical aspects for LHC operation and commissioning. In particular, it must be ensured that the triplet magnets in the experiments are protected during the betatron squeeze [3].

- Active cleaning of collision debris products: this is achieved with dedicated (TCL) collimators located on the outgoing beams of each high luminosity experiment, which catch the debris produced by the collisions. These collimators keep losses below the quench limit of the superconducting magnets in the matching sections and dispersion suppressors close to the interaction points.

- Experiment background optimization: this is one of the classical roles of collimation systems in previous colliders like the ISR, the SppS, and the Tevatron. For the LHC, the contribution to background from beam halo has always been expected to be small, due to effective IR7 collimation cleaning that induces only limited losses close to the experiments. The initial run confirmed this expectation [4].

- Concentration of radiation losses: for high power machines, it is becoming increasingly important to be able to localize beam losses in confined and optimized 'hot' areas rather than having a distributed activation of equipment along the machine. This is an essential functionality to allow easy access for maintenance in the largest parts of the machine.

- Local protection of equipment and improvement of lifetime: dedicated movable or fixed collimators are used to shield equipment. For example, eight passive absorbers are used in the collimation insertions in order to reduce the total dose to warm dipoles and quadrupoles that otherwise would have a short lifetime in the high-radiation environment foreseen during nominal LHC operation.

- Beam halo scraping and halo diagnostics: collimator scans in association with the very sensitive LHC beam loss monitoring system proved to be a powerful way to probe the population of beam tails [5, 6], which were otherwise too small compared to the beam core to be measured by conventional emittance measurements. Thanks to their robustness, the present primary collimators can also be efficiently used to scrape and shape the beams, as in Ref. [7].

In order to fulfil all these functionalities, the LHC collimation system features an unprecedented complexity compared to the previous state-of-the-art in particle accelerators. The Run 1 system required managing about 400° of freedom for collimator movements [8]. As a comparison, the Tevatron collimation system had less than 30° of freedom. For this reason, the possibility of reliably operating the collimation system has always been considered to be a major concern for LHC performance. Upgrade scenarios must address improved operational aspects, as the HL-LHC goal relies on machine availability.

Table 5-1: Collimators for the LHC Run 2, starting in 2015. For each type, acronyms, rotation plane (horizontal, vertical or skew), material and number of devices, summed over the two beams, are given. For completeness, movable injection and dump protection devices are also listed. In addition, the collimation system comprises 10 fixed-aperture absorbers in IR3 and IR7 to reduce total doses to worm magnets of the cleaning insertions.

| Functional type | Name | Plane | Number | Material |
|---|---|---|---|---|
| Primary IR3 | TCP | H | 2 | CFC |
| Secondary IR3 | TCSG | H | 8 | CFC |
| Absorber IR3 | TCLA | H, V | 8 | Inermet 180 |
| Primary IR7 | TCP | H, V, S | 6 | CFC |
| Secondary IR7 | TCSG | H, V, S | 22 | CFC |
| Absorber IR7 | TCLA | H, V, S | 10 | Inermet 180 |
| Tertiary IR1/IR2/IR5/IR8 | TCTP | H, V | 16 | Inermet 180 |
| Physics debris absorbers IR1/IR5 | TCL | H | 12 | Cu, Inermet180 |
| Dump protection IR6 | TCDQ | H | 2 | CFC |
| Dump protection IR6 | TCSP | H | 2 | CFC |
| Injection protection (transfer lines) | TCDI | H, V | 13 | C |
| Injection protection IR2/IR8 | TDI | V | 2 | hBN, Al, Cu/Be |
| Injection protection IR2/IR8 | TCLI | V | 4 | C, CFC |
| Injection protection IR2/IR8 | TCDD | V | 1 | Copper |

## 5.2 Baseline upgrades to the LHC collimation system

To cope with the increased challenges in the HL-LHC, several of the functionalities of the LHC collimation system must be upgraded. We discuss how to improve the cleaning performance, the impedance, and the collimation in the experimental IRs.

5.2.1   Upgrades for cleaning improvement

**5.2.1.1   Upgrades of the IR7 system**

Protons and ions interacting with the collimators in IR7 emerge from the IR with a modified magnetic rigidity. This represents a source of local heat deposition in the cold dispersion suppressor (DS) magnets downstream of IR7, where the dispersion starts to increase (see Ref. [16] and references therein): these losses are the highest cold losses around the ring. This may pose a certain risk for inducing magnet quenches, in particular in view of the higher intensities expected for the HL-LHC.

A possible solution to this problem is to add local collimators in the dispersion suppressors, which is only feasible with a major change of the cold layout at the locations where the dispersion start rising. Indeed, the present system's multi-stage cleaning is not efficient at catching these dispersive losses. Clearly, the need for local collimation depends on the absolute level of losses achieved in operation and the quench limit of superconducting magnets. In this design phase, where the quench limits and the operational performance are not yet known accurately enough at beam energies close to 7 TeV, it is important to take appropriate margins to minimize the risk of being limited in the future (LHC operation at beam intensity above nominal design, and even more in the HL-LHC era).

A strategy to eliminate any risk of quench is the installation of DS collimators (target collimator long dispersion suppressors (TCLDs)). As shown below, two collimators per side of IR7, one per beam on each side, would be sufficient to effectively intercept the protons or ions that would otherwise hit the DS magnets. In order to make space for the new collimators, it is envisaged to replace, for each TCLD, an existing main dipole with two shorter 11 T dipoles with the TCLD in between, as shown in Figure 5-5. This is a modular solution that can actually be applied to any dipole without additional changes to the adjacent superconducting magnets or other cold elements, should a space in the continuous cryostat be needed for any reason in the future [17].

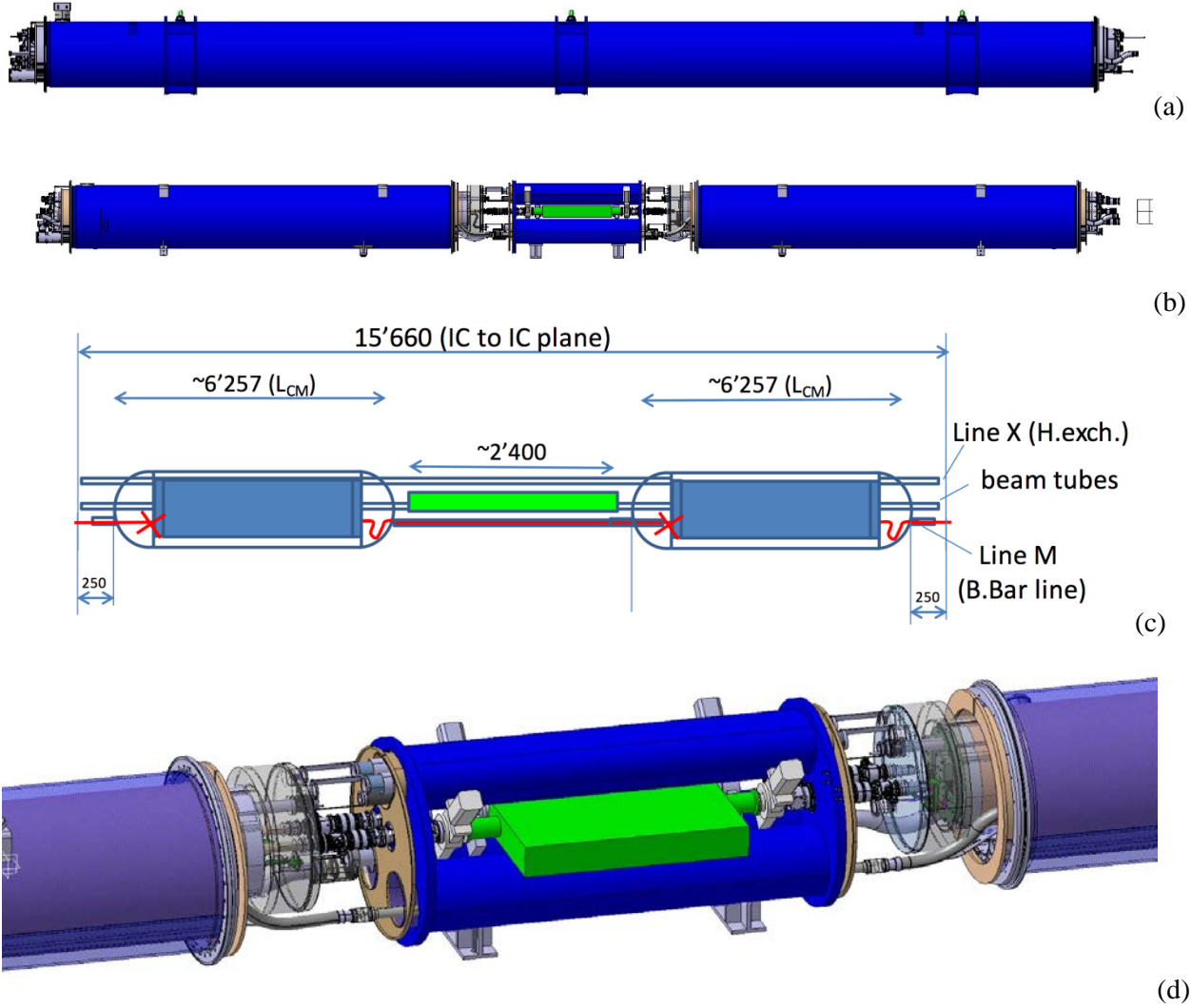

Figure 5-5: (a–c) Schematic view of the assembly of two shorter 11 T dipoles with a collimator in between, which can replace one standard main dipole. (Courtesy of V. Parma.) (d) Preliminary 3D model of a TCLD assembly showing the collimator (in green), the two short dipole cryostats and the connection cryostat. Note the very tight space constraints for the collimator unit.

Extensive tracking and energy deposition simulations have been performed to assess the effect of the TCLDs [18–22], based on the assumption that the dipoles MB.B8R7 and MB.B10R7 are substituted for cleaning B1, and MB.B8L7 and MB.B10L7 for cleaning B2. This layout makes room for two TCLDs per beam. For example, the simulated energy deposition profile of the DS magnets for the case of 0.2 h lifetime in the nominal LHC beam is illustrated in Figure 5-6. It can be seen that the presence of local DS collimators reduces the peak energy deposition by about a factor of 10 compared to the present layout with standard dipoles.

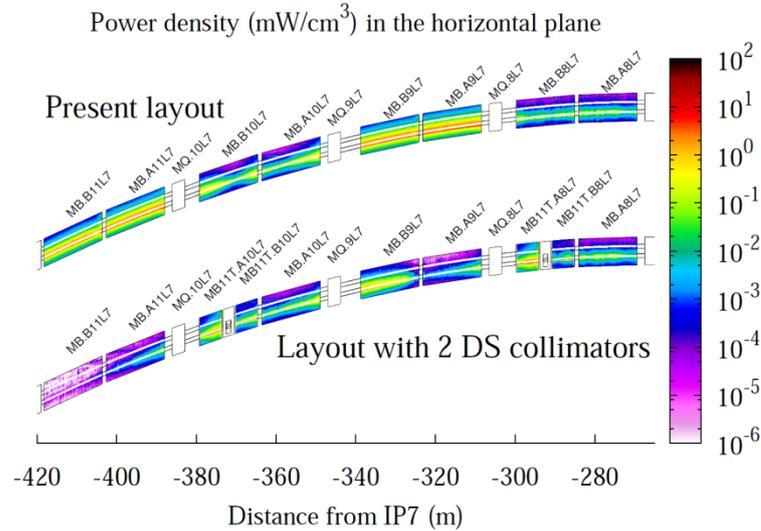

Figure 5-6: Simulated power density map in the horizontal plane of DS dipoles for nominal 7 TeV operation and a beam lifetime of 0.2 h ($4.5 \times 10^{11}$ protons lost per second). The map makes a comparison between the present layout and a layout with two TCLDs. Results correspond to relaxed collimator settings. Beam direction is from the right to the left. From Ref. [19].

TCLD collimators also make the cleaning performance more robust against various errors of the collimation system, of the optics, and of the orbit [22], as they remove the off-momentum particles at the first high-dispersion location downstream of IR7. This is of particular concern for the ATS optics, which require modified optics in the cold arcs. Indeed, for the HL baseline optics, this solution almost eliminates losses around the ring coming from the telescopic squeeze. Should the total intensity be limited by collimation cleaning, the factor of 10 quoted above would translate into the same gain factor for the total stored beam energy.

Furthermore, the improvement in cleaning could be very beneficial for LHC operation even if this is not limited by the collimation losses. For example, a better cleaning performance might allow relaxation of the opening of some secondary collimators with a subsequent reduction of machine impedance. It should also be noted that the DS collimation solution might also mitigate issues related to radiation damage to cold magnets protected by the TCLD collimators and the activation of near-by components.

The real need for this gain can only be addressed after having accumulated beam experience at higher energies during post-LS1 operation (including beam tests of quench limits at energies close to 7 TeV). On the other hand, a recent collimation project review recommended that the preparation of DS collimation in IR7 be pursued with a high priority [23].

Even if the full performance improvement provided by the DS collimation solution in IR7 relies on two TCLD collimators per side, alternative solutions based on one single unit are being considered for possible 'staged' deployment in IR7, in case performance limitations during high-intensity proton operation are made apparent by the post-LS1 operation experience.

The TCLD collimators' design is derived from one of the standard LHC collimators [24]. In particular, it incorporates the latest design improvements, such as in-jaw BPMs. TCLD collimators require a reduced set of control cables because each jaw will be moved by a single motor; though they still require cooling water and baking equipment. The key parameters are listed in Table 5-2. Although some design features are less demanding, which is also due to lower losses compared to other collimators, integration design aspects are much more complicated due to their location between cold elements.

Table 5-2: Key parameters of TCLD collimators

| Characteristics | Units | Value |
|---|---|---|
| Jaw active length | [mm] | 80 |
| Jaw material | - | Inermet 180 |
| Flange-to-flange distance | [mm] | 1080 |
| Number of jaws | - | Two |
| Orientation | - | Horizontal |
| Dipole replaced by 11 T dipole/TCLD | - | MB.B10 |
| Number of BPMs per jaw | - | Two |
| RF damping | - | RF fingers or ferrite |
| Cooling of the jaw | - | Yes |
| Cooling of the vacuum tank | - | No |
| Minimum gap | [mm] | <2 |
| Maximum gap | [mm] | >45 |
| Stroke across zero | [mm] | >4 |
| Number of motors per jaw | - | One |
| Angular adjustment | - | No |
| Transverse jaw movement (fifth axis) | - | No |

The new baseline that relies on shorter 11 T dipoles has been reviewed from the integration point of view [25]. Space is restricted, and the length of all components and transitions must be carefully optimized. The present baseline is that the TCLD will have an active jaw length of 80 cm, which has proved to be sufficient to improve cleaning in all relevant cases. Tungsten heavy alloy is assumed for the material because the TCLD will rarely be exposed to a large beam load, so there is no need at this stage to consider advanced materials. From the RF view point, designs with transverse RF fingers (as in the present system) as well as with ferrite blocks to absorb high-order modes (as in the collimators with BPMs) are being comparatively assessed. A possible design is shown in Figure 5-7, where a detail of the collimator jaw extremity is given.

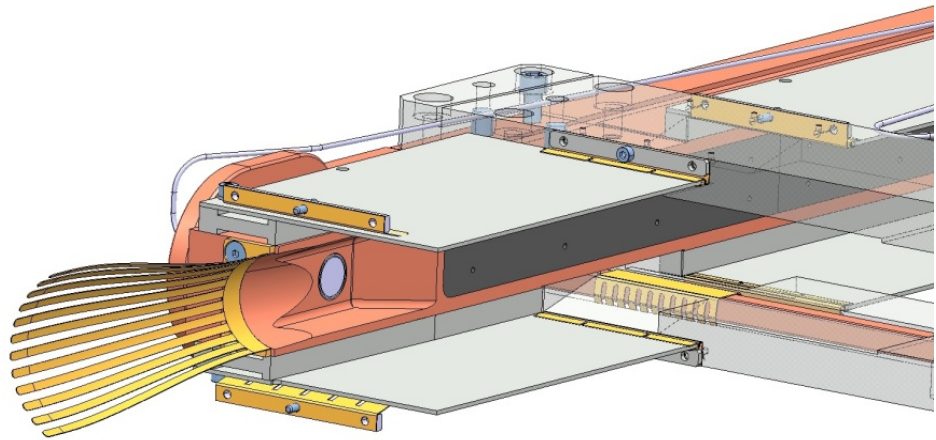

Figure 5-7: Detail of one corner of the TCLD collimator to be installed in the DS between two new 11 T dipoles. The present design foresees an 80 cm long jaw made of tungsten (the first of four 20 cm tungsten tiles is shown) and will have two jaws. Designs with transverse RF fingers or ferrite tiles are being comparatively assessed to reduce the detrimental effects of trapped RF modes.

### 5.2.1.2 Upgrades for improved cleaning of physics debris close to experiments

Collision products emerging from the interaction points might be lost in the matching sections and the dispersion suppressors (DS) around the experiments. In particular, protons that changed their magnetic rigidity represent a source of local heat deposition in the first DS cells where the dispersion function starts rising. These physics debris losses may pose a certain risk of inducing magnet quenches.

Mechanisms with similar effects also occur during heavy-ion operation [25–28]. Secondary ion beams with a changed magnetic rigidity are created when ions undergo ultra-peripheral interactions at the collisions. The dominating processes are bound-free pair production (BFPP), where electron–positron pairs are created and an electron is caught in a bound state by one (BFPP1) or both (BFPP2) nuclei, thus changing their charge, and 1- or 2-neutron electromagnetic dissociation (EMD1 and EMD2) where one of the colliding ions emits one or two neutrons, respectively, thus changing mass. Further photo-induced processes also take place, but the four mentioned here have the higher cross-sections. An example of ion beams produced in collisions of $^{208}$Pb$^{82+}$ nuclei in IR2 is given in Figure 5-8.

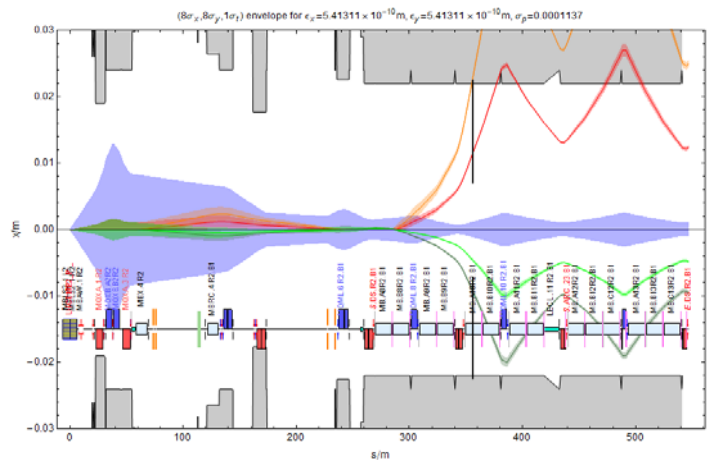

Figure 5-8: 1 $\sigma$ envelope of the main Pb$^{82+}$ beam (violet) together with the dispersive trajectories of ions undergoing BFPP1 (red), BFPP2 (orange), EMD1 (light green), and EMD2 (dark green), coming out of the ALICE experiment in nominal optics. The DS collimator appears as a black line. Varying its opening allows different secondary beams to be intercepted (note that the orange BFPP2 beam carries energies well below the quench limit).

As can be seen, these secondary beams are lost very locally due to the big and sudden change of magnetic rigidity, and may pose a risk of inducing magnet quenches [16, 28]. For the LS2 ALICE upgrade, aiming at a peak luminosity of $6 \times 10^{27}$ cm$^{-2}$ s$^{-1}$ (about six times higher than the nominal one), the dominant BFPP1 beam can carry about 150 W, resulting in a power load in the coils of the MB.B10 dipole of about 50 mW/cm$^3$ [29], in the DS regions on both sides of ALICE. This is a factor of about 2 above the quench limit according to figures presented at a recent collimation review [23]. Similar losses also occur in the DS regions around ATLAS and CMS during ion operation.

A strategy to eliminate any risk of quenches in the experimental IRs, both for proton and heavy-ion runs, is the installation of TCLD collimators in the DS to catch ions beams before they reach the magnets, as shown schematically in Figure 5-8. For heavy-ion operation, one collimator per side of the experiment would be sufficient to effectively intercept the secondary beams from the most dominant processes in a location where these ions are well separated from the main beam.

Extensive energy deposition simulations in the DS around ALICE, where MB.A10 is substituted by a pair of 11 T magnets and a TCLD collimator, confirm this assumption [16, 29]. The proposed TCLD collimator installation, with a jaw based on 80 cm of a tungsten heavy alloy, is expected to reduce by more than a factor of 100 the peak power density in the new 11 T dipoles compared to the power density in cold dipoles with the present layout with old dipoles and no TCLD collimators [29]. DS collimation around IR2 is therefore considered to be necessary for the full exploitation of the ALICE detector upgrade.

Different jaw lengths and materials have been comparatively addressed for the specific case of ions in IR2 by using, as an advantageous comparison, the reduction factor of losses in the DS dipoles [29]. Simulations show that 50 cm of copper would suffice (see, for example, the results presented at the collimation review [23]). However, in order to minimize design effort and production works both for the collimator and for the design of the cryo bypass, the same length of 80 cm adopted for the TCLD in IR7 (see Table 5-2) is also used as a baseline for the TCLD collimators in the experimental IRs. For the jaw material, the tungsten alloy Inermet 180 is used as a baseline for IR7. Should the copper design be easier/less costly, it could be considered for implementation around P2.

For high intensity proton operation, the losses observed around IR1 and IR5 can be reduced with two TCLD collimators per IR side. The need for such implementation depends on the dipole quench limits and on the effectiveness of the physics debris collimation with TCL collimators. Layouts based on TCL collimators only (positioned in the straight section of IR1 and IR5 where the local dispersion from the collision point is low) might be sufficient but this requires further studies with the latest HL-LHC layouts that will estimate the peak energy deposited in the DS magnets. If required the TCLDs would then complement the present system with TCLs.

### 5.2.2  Upgrades for impedance improvement

The LHC impedance budget is largely dominated by the contribution of the LHC collimators. For this reason, the present collimation system has been conceived in a way that it can be easily upgraded to reduce the impedance [30]: every secondary collimator slot in IR3 and IR7 features a companion slot for the future installation of a low-impedance secondary collimator. A total of 22 slots (IR7) and eight slots (IR3) are already cabled for a quick installation of new collimators – referred to as TCSMP in the present database naming convention – that can either replace the present TCSGs or be used together with them. Partial preparation of these slots is ongoing in LS1.

The importance of minimizing the machine impedance for the HL-LHC has been emphasized in Refs. [31–33] and also in a recent LHC collimation review [23]. We therefore foresee that, by the time of the full HL-LHC implementation (LS3), some or all of the available TCSMP slots might be equipped with advanced collimators using new materials, and possibly coatings, to reduce the impedance. A staged installation using the various technical stops and shutdowns after LS1 is possible according to actual needs.

Secondary collimators in the betatron cleaning insertion (IR7) also have a crucial role in LHC machine protection and might be exposed to large beam losses. Therefore, new material choices and designs must also be robust against beam failure (at the least those exposed to horizontal losses). The driving requirements for the development of new materials are thus: (i) low resistive-wall impedance to avoid beam instabilities; (ii) high cleaning efficiency; (iii) high geometrical stability to maintain the extreme precision of the collimator jaw during operation despite temperature changes; and (iv) high structural robustness in case of accidental events like single-turn losses.

The present baseline for the upgraded secondary collimators relies thus on molybdenum carbide-graphite (MoGr) composites, possibly coated with pure molybdenum (Figure 5–9). Other lower $Z$ refractory coatings are presently under study. The molybdenum coating under consideration would reduce the surface resistivity by about a factor of 10 to 20 compared to Mo-Gr and by more than a factor of 100 compared to CFC. The benefit on the impedance budget of the collimation system would be significant: in the relevant frequency range, impedance would be reduced to 10% of that of the CFC jaws [34], as illustrated in Figure 5-10.

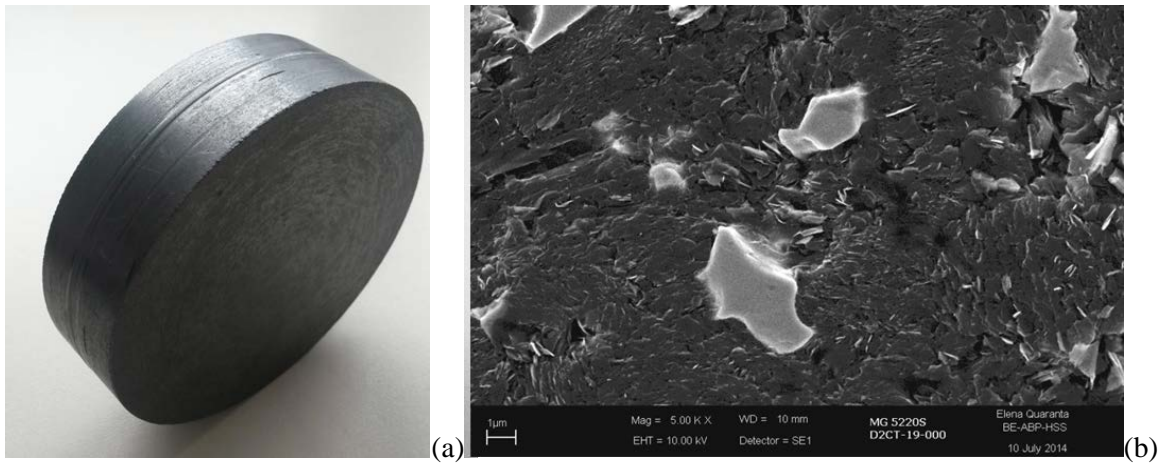

Figure 5-9: (a) Mo-Gr plate recently produced by Brevetti Bizz, Italy. Dimensions of the plate: 90 mm diameter and 24.3 mm thickness. It is a massive piece prepared in view of the production of the LHC collimator jaw inserts. (b) A detail of the microstructure, where the graphite flakes matrix well sintered with the carbon fibers is visible together with a few molybdenum carbide 'islands' of about 5 μm length.

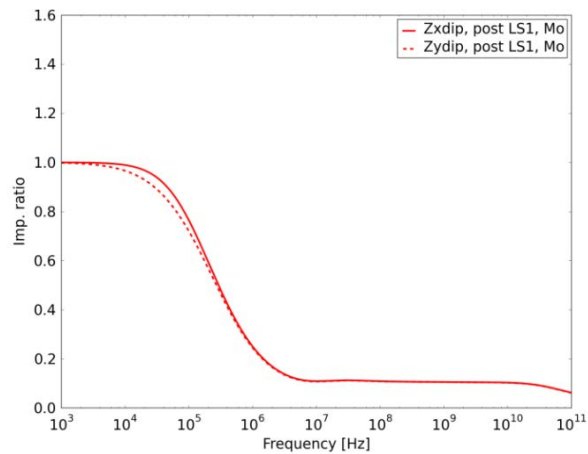

Figure 5-10: Collimation impedance versus frequency: impedance ratio between Mo coating on Mo-Gr (50 μm layer) and present CFC jaw for the real (solid) and imaginary (dotted) parts. A secondary collimator is considered. Courtesy of N. Mounet.

On the other hand, the new design and materials [35] must be validated for operation. Material properties and the coating options have to be validated for operation in the LHC. For these purposes, a rich programme of validation is in progress, involving:

- tests at HiRadMat, covering both material samples as well as full jaw validation;
- mechanical engineering prototyping;
- beam tests at the LHC, planned for 2016 (collimation installation in the 2015 shutdown).

In addition to the impedance improvements, the new TCSPM also feature a number of improvements in the mechanical design (Figure 5-11) [35]. They incorporate the BPM button design. The key hardware parameters are listed in Table 5-3.

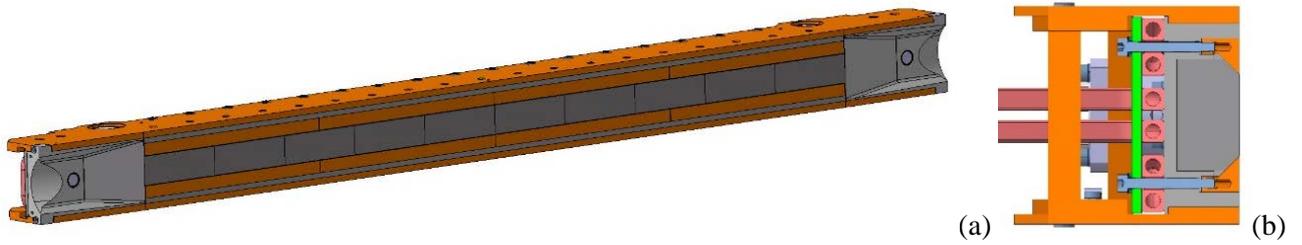

(a)  (b)

Figure 5-11: Preliminary design of the TCSMP jaw (a) and of its cross-section (b). The jaw assembly features 10 MoGr blocks. Also note that the jaw tapering is lengthened, further reducing its contribution to HOM RF instabilities in the geometrical transition zones

Table 5-3: Parameters of TCSMP collimators

| Characteristics | Units | Value |
|---|---|---|
| Jaw active length | [mm] | 1000 |
| Jaw material | - | MoGr |
| Flange-to-flange distance | [mm] | 1480 |
| Number of jaws | - | 2 |
| Orientation | - | Horizontal, vertical, skew |
| Number of motors per jaw | - | Two |
| Number of BPMs per jaw | - | Two |
| RF damping | - | Fingers |
| Cooling of the jaw | - | Yes |
| Cooling of the vacuum tank | - | Yes |
| Minimum gap | [mm] | <1 |
| Maximum gap | [mm] | 60 |
| Stroke across zero | [mm] | 5 |
| Angular adjustment | - | Yes |
| Jaw coating | - | Mo (to be confirmed) |
| Transverse jaw movement (fifth axis) | [mm] | ±10 |

### 5.2.3 Upgrades to the collimation of the incoming beam in the experimental IRs

The LHC Run 1 operation period has shown that protection of the IR superconducting magnets and experiments is a key asset for machine performance: the available aperture, to be protected in all operational phases, determines the collimation hierarchy. The present tertiary collimators (target collimator tertiary with pick-up (TCTP)) are located at positions that protect the triplet and are made of a heavy tungsten alloy (Inermet 180). They effectively protect the elements downstream but are not robust against high beam losses, in particular during very fast beam failures that might occur if the beam dumping system does not trigger synchronously with the abort gap (an asynchronous beam dump). Settings margins are added to the collimator hierarchy to minimize the risk of exposure of these collimators to beam losses in case of such failures [3]. A design with improved robustness would allow the reduction of these margins and, as a result, push further the $\beta^*$ performance of the LHC, in particular for the HL optics baseline (ATS) that features an unfavourable phase between dump kickers and triplet magnets.

In addition to improvements from increased robustness, the HL-LHC layout has additional aperture constraints [2, 3] because the aperture of the magnets up to Q5 is now smaller than in the present layout. Thus, up to four more tertiary collimators might be required in IR1/IR5 to protect the Q4 and Q5 quadrupole magnets, in addition to those installed to protect the triplet (two TCTP collimators – one horizontal and one vertical). The present baseline under study includes also a pair of new collimators in front of Q5. Ongoing studies are addressing: (i) the need for additional Q4 protection; and (ii) the need to keep tertiary collimators at the present locations in case additional tertiaries are added upstream.

A new design of tertiary collimators, referred to as target collimator tertiary with pick-up metallic (TCTPM), is under study to address the new challenges. This design will be based on novel materials to improve collimator robustness while ensuring adequate absorption, adequate cleaning, and protection of the elements downstream. The TCTPM design and material choice must also take impedance constraints under consideration to keep the collimator impedance under control. A summary of the technical key parameters are given in Table 5-4.

The experimental experience of beam impacts on collimator material samples at HiRadMat [4, 5] indicates that a molybdenum-graphite (Mo-Gr) composite can improve the TCTP robustness by a factor of several hundreds. Note that the present Inermet design is expected to undergo severe damage requiring a collimator replacement if hit by one single LHC nominal bunch of $10^{11}$ proton at 7 TeV. Other advanced materials are being studied as possible alternatives to further improve the robustness. The HL beam parameters with bigger charge and smaller emittance pose additional challenges in terms of beam damage potential.

Despite being less critical because of the larger $\beta^*$ values, upgraded TCTP's are under consideration also for IR2/8 because various luminosity scenarios in there IRs require the usage of tertiary collimators, although at relaxed settings compared to IR1 and IR5.

Table 5-4: Equipment parameters of the TCTPM

| Characteristics | Units | Value |
|---|---|---|
| Jaw active length | [mm] | 1000 |
| Jaw material | - | Mo-Gr (to be decided) |
| Flange-to-flange distance | [mm] | 1480 (to be reviewed) |
| Number of jaws | - | Two |
| Orientation | - | Horizontal, vertical |
| Number of motors per jaw | - | Two |
| Number of BPMs per jaw | - | Two |
| RF damping | - | Fingers |
| Cooling of the jaw | - | Yes |
| Cooling of the vacuum tank | - | Yes |
| Minimum gap | [mm] | <1 |
| Maximum gap | [mm] | >60 (to be reviewed) |
| Stroke across zero | [mm] | >5 |
| Angular adjustment | - | Yes |
| Jaw coating | - | No |
| Transverse jaw movement (fifth axis) | [mm] | ±10 mm (at least) |

## 5.3 Advanced collimation concepts

In this section we discuss new, more advanced, collimation concepts that still require R&D and are therefore not in the baseline. However, depending on the results of Run 2, some of these concepts may become an important asset for the LHC and the HL-LHC.

### 5.3.1 Halo diffusion control techniques

The 2012 operational experience indicates that the LHC collimation would profit from halo control mechanisms. These mechanisms were used in other machines like HERA and Tevatron. The idea is that, by controlling the diffusion speed of halo particles, one can act on the time profile of the losses, for example by reducing rates of losses that would otherwise take place in a short time, or simply by controlling the static population of halo particles in a certain aperture range. These aspects were recently discussed at a collimation review on the possible usage of the hollow e-lens collimation concept at the LHC [36], where it was concluded that hollow e-lenses could be used at the LHC for this purpose. In this case, a hollow electron beam runs parallel to the proton or ion beam that is on the axis of the cylindrical layer of electron. This hollow beam produces an electromagnetic field only affecting halo particles above a given transverse amplitude, changing their transverse speed. The conceptual working principle is illustrated in Figure 5-12(a). A solid experimental basis achieved at the Tevatron indicates this solution is promising for the LHC ([37] and reference therein).

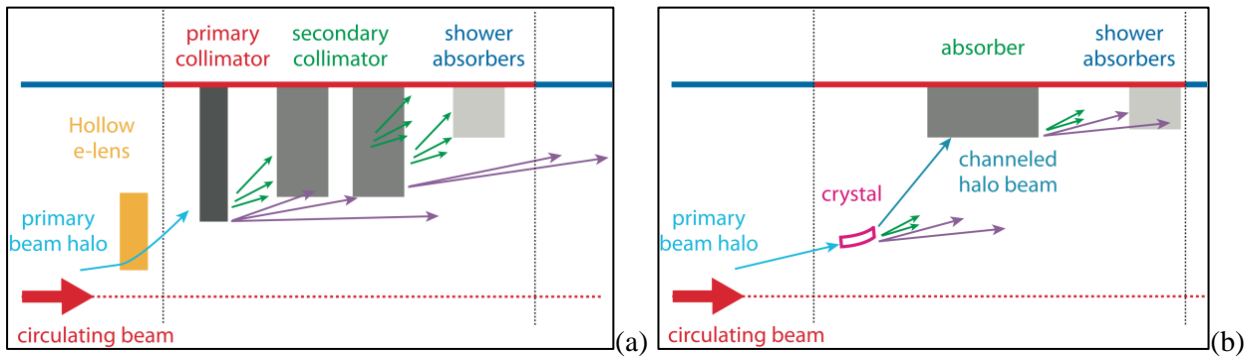

Figure 5-12: Illustrative view (a) of the collimation system with integrated hollow e-lens or equivalent halo diffusion mechanism; (b) an ideal crystal-based collimation. A simplified collimator layout to that in Figure 5-1 is adopted to show the betatron cleaning functionality only (one side only). Halo control techniques are used to globally change the diffusion speed of halo particles, and rely on the full collimation system remaining in place. Crystals entail a change of concept where the whole beam losses are concentrated, ideally, in one single beam absorber per plane.

The potential advantages of the electron lens collimation are several.

- Control of the primary loss rates, with potential mitigation of peak loss rates in the cold magnets, for a given collimation cleaning. Peak power losses on the collimators themselves can be optimized as well.
- Controlled depletion of beam tails, with beneficial effects in case of fast failures.
- Reduction of tail populations and therefore peak loss rates in the case of orbit drifts.
- Beam scraping at very low amplitudes (>3 σ) without the risk of damage, as for bulk scrapers.
- Tuning of the impact parameters on the primary collimators with a possible improvement in cleaning efficiency.

Since the main beam core is not affected, hollow electron beam (HEB) operation should in theory be transparent for the luminosity performance if this technique works as designed. This was demonstrated at the Tevatron for DC powering of the electron beams.

The use of HEB requires the collimation system to be in place in order to dispose of the tail particles expelled in a controlled way. No losses occur at the HEB location and the tail control mechanism can be put in place in any ring location. Larger beam size locations are favourable as they entail reduced alignment accuracy for the hollow beam. IR4 is considered to be the best candidate for two HEB devices due to the larger than standard inter-beam distance (which eases integration of the device on the beam), cryogenics availability, low-radiation environment, and quasi-round beam.

While the functionality of HEB will provide clear benefits for LHC operation, the real need for such a scheme at the LHC and the HL-LHC has to be addressed after gaining sufficient operational experience at energies close to 7 TeV on quench limits, beam lifetime, and loss rates during the operational cycle and collimation cleaning. Fast failure scenarios for the crab cavities require a low tail population above about 4 beam $\sigma$: HEB is the only technique solidly validated experimentally in other machines that could in this case ensure safe operation.

The HEB is targeted at enabling active control of beam tails above 3 beam $\sigma$, with tail depletion efficiencies of the order of 90% over times of tens of seconds, in all phases of the operational cycle, specifically before and after beams are put into collision.

The HEB implementation should ensure: (i) the possibility of pulsing the current turn-by-turn (as required to drive resonances in the linear machine before beams are in collision); (ii) a train-by-train selective excitation (leaving 'witness' trains with populated halos for diagnostics and machine protection purposes).

The main systems/components of a HEB can be summarized as:

- electron beam generation and disposal: electron gun and collector, with the required powering;
- several superconducting and resistive magnets: solenoids, dipoles, and correctors to stabilize and steer the electron beam;
- beam instrumentation for the optimization of the electron beam.

The parameters listed here are extracted from the conceptual design document [38] compiled by colleagues from FNAL who worked on this topics within the LARP collaboration. A detail engineering design is now ongoing at CERN. The first goal will be to define the volumes for a full integration into the LHC.

Table 5-5: Hollow electron beam equipment parameters

| Parameter | Value or range |
|---|---|
| *Beam and lattice* | |
| Proton kinetic energy, $T_p$ [TeV] | 7 |
| Proton emittance (rms, normalized), $\varepsilon_p$ [$\mu$m] | 3.75 |
| Amplitude function at electron lens, $\beta_{x,y}$ [m] | 200 |
| Dispersion at electron lens, $D_{x,y}$ [m] | $\leq 1$ |
| Proton beam size at electron lens, $\sigma_p$ [mm] | 0.32 |
| *Geometry* | |
| Length of the interaction region, $L$ [m] | 3 |
| Desired range of scraping positions, $r_{mi}$ [$\sigma_p$] | 4–8 |
| *Magnetic fields* | |
| Gun solenoid (resistive), $B_g$ [T] | 0.2–0.4 |
| Main solenoid (superconducting), $B_m$ [T] | 2–6 |
| Collector solenoid (resistive), $B_c$ [T] | 0.2–0.4 |
| Compression factor, $k \equiv \sqrt{B_m/B_g}$ | 2.2–5.5 |
| *Electron gun* | |
| Inner cathode radius, $r_{gi}$ [mm] | 6.75 |
| Outer cathode radius, $r_{go}$ [mm] | 12.7 |
| Gun perveance, $P$ [$\mu$perv] | 5 |
| Peak yield at 10 kV, $I_e$ [A] | 5 |
| *High-voltage modulator* | |
| Cathode-anode voltage, $V_{ca}$ [kV] | 10 |
| Rise time (10%–90%), $\tau_{mod}$ [ns] | 200 |
| Repetition rate, $f_{mod}$ [kHz] | 35 |

Source: Ref. [38]

At the collimation review [36], it became clear that, if loss spikes were limiting LHC performance after LS1, the hollow e-lens solution would not be viable because it could only be implemented in the next long shutdown at the earliest (driven by the time for integration into the cryogenics system). It is therefore crucial to work on viable alternatives that, if needed, might be implemented in an appropriate time scale. Two alternatives are presently being considered:

- tune modulation through a ripple in the current of lattice quadrupoles;
- narrow-band excitation of halo particles with the transverse damper system.

Though very different from the hardware point of view, both these techniques rely on exciting tail particles through resonances induced in the tune space by appropriate excitations. This works on the assumption of the presence of a well-known and stable correlation between halo particles with large amplitudes and corresponding tune shift in tune space (de-tuning with amplitude). Clearly, both methods require a solid experimental verification in a very low noise machine like the LHC, in particular to demonstrate that this type of excitation does not perturb the beam core emittance. Unlike hollow e-lenses, which act directly in the transverse plane by affecting particles at amplitudes above the inner radius of the hollow beam, resonance excitation methods require a good knowledge of the beam core tune even in dynamic phases of the operational cycle, so the possibility of using these techniques at the LHC remains to be demonstrated. For this purpose, simulation efforts are ongoing with the aim of defining the required hardware interventions during LS1 that might enable beam tests of these two halo control methods early on in 2015. Ideally, these measurements would profit from appropriate halo diagnostic tools. We are, however, confident that conclusive measurements could be achieved in Run 2 with the techniques described, for example in Ref. [5].

### 5.3.2 Crystal collimation

Highly pure bent crystal can be used to steer high-energy particles that get trapped by the potential of parallel lattice planes. Equivalent bending fields of up to hundreds of tesla can be achieved in crystals with a length of only 3–4 mm, which allows in principle steering of halo particles to a well-defined point. As opposed to a standard collimation system based on amorphous materials, requiring several secondary collimators and absorbers to catch the products developed through the interaction with matter (Figure 5-1), one single absorber per collimation plane is in theory sufficient in a crystal-based collimation system [39]. This is shown in the scheme in Figure 5-12(b). Indeed, nuclear interactions with well-aligned crystals are much reduced compared to a primary collimator, provided that high channelling efficiencies for halo particles can be achieved (particles impinging on the crystal to be channelled within a few turns). This is expected to significantly reduce the dispersive beam losses in the DS of the betatron cleaning insertion (IR7) compared to the present system, which is limited by the leakage of particles from the primary collimators. Simulations indicate a possible gain of between 5 and 10 [40], even for a layout without an optimized absorber design. The crystal collimation option is particularly interesting for collimating heavy-ion beams thanks to the reduced probability of ion dissociation and fragmentation compared to the present primary collimators. SPS test results are promising [41].

Another potential advantage of crystal collimation is a strong reduction of machine impedance due to the facts that: (i) only a small number of collimator absorbers are required; and that (ii) the absorbers can be set at much larger gaps thanks to the large bending angle from the crystal (40–50 μrad instead of a few μrad from the multiple-Coulomb scattering in the primary collimator). On the other hand, an appropriate absorber design must be conceived in order to handle the peak loss rates in case of beam instabilities: the absorber must withstand continuous losses of up to 1 MW during 10 s while ensuring the correct collimation functionality. This is a change of paradigm compared to the present system, where such losses are distributed among several collimators. Other potential issues concern the machine protection aspects of this system (such as the implications of a crystal not being properly aligned and therefore channelling an large fraction of the total stored energy to the wrong place) and the operability of a system that requires mechanical angular stability in the sub-μrad range to be ensured through the operational cycle of the LHC (injection, ramp, squeeze, and collision).

Promising results have been achieved in dedicated crystal collimation tests at the SPS performed from 2009 within the UA9 experiment [41–43]. On the other hand, some outstanding issues about the feasibility of the crystal collimation concept for the LHC can only be addressed by dedicated beam tests at high energy in the LHC. For this purpose, a study at the LHC has been proposed [44]: two goniometers housing crystals have been installed in IR7 during LS1 for horizontal and vertical crystal collimation tests. The main purpose of beam tests at the LHC is to demonstrate the feasibility of the crystal-collimation concept in the LHC environment, in particular to demonstrate that such a system can provide a better cleaning of the present high-performance system throughout the operational cycle. Until a solid demonstration is achieved, crystal collimation schemes cannot be considered for future HL-LHC baseline scenarios.

### 5.3.3 Improved optics scenarios for collimation insertions

Alternative optics concepts in IR7 can be conceived in order to improve some present collimation limitations without major hardware changes. For example, non-linear optics schemes derived from the linear collider experience [45] were considered for IR7. The idea is that one can create a 'non-linear bump' that deforms the trajectories of halo particles and effectively increases their transverse amplitudes in a way that allows opening the gaps of primary and secondary collimators. These studies are well advanced from the optics point of view but at the present time it is not possible to easily find a layout solution providing the same cleaning as the present system [46]. These studies, and others aimed at increasing the beta functions at the collimators, are ongoing.

### 5.3.4 Rotatory collimator design

The rotatory collimator design developed at SLAC under the LARP effort proposes a 'consumable collimator' concept based on two round jaws with 20 flat facets that can be rotated to offer to the beam a fresh collimator material in case a facet is damaged [47]. This design features a low impedance and is based on standard non-exotic materials. It was conceived for high-power operation, with a 12 kW active cooling system to withstand the extreme power loads experienced by the secondary collimators in IR7. A photograph of this device before closing the vacuum tank is given in Figure 5-13, where the rotatory glidcop (a copper alloy) jaws are visible. The first full-scale prototype of this advanced collimator concept has recently been delivered to CERN [48] and is being tested in preparation for beam tests. The ultimate goal is to validate the rotation mechanism after high intensity shock impacts at the HiRadMat facility, aimed at demonstrating that the concept of consumable collimator surfaces can indeed work for the LHC beam load scenarios. The precision accuracy of this prototype and the impedance are also being tested together with its vacuum performance. The vacuum measurements indicated that the SLAC prototype is suitable for installation in the SPS or even the LHC. An optimum strategy for beam tests is being established based on these new results.

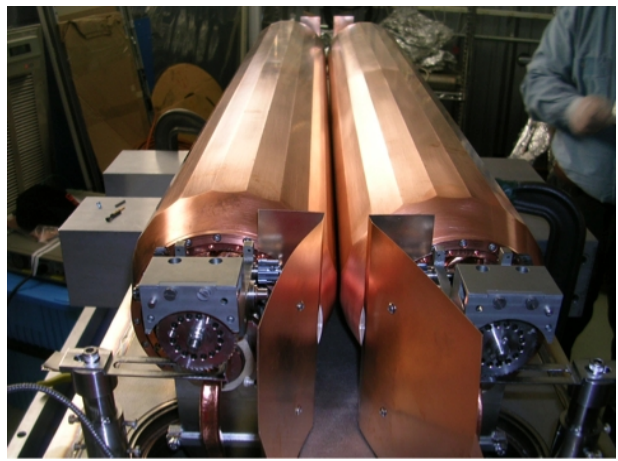

Figure 5-13: Photograph of the SLAC rotatory collimator prototype jaws before assembly in the vacuum tank. Courtesy of T. Markiewicz (SLAC, LARP).

## 5.4 Other collimators from the present system required in the HL-LHC

It is important to realize that more than 60% of the LHC collimators, which are not to be modified or replaced in the present collimation upgrade baseline described above, must remain reliably operational for the HL-LHC era. Even devices whose design is deemed adequate for the HL-LHC parameters can hardly survive for the lifetime of the LHC machine without appropriate maintenance, substitution, or revamping. A long-term strategy must be put in place in order to ensure that the LHC collimation system can meet the performance and availability challenges of the HL-LHC project. In this section, the present LHC collimators that will also be needed for the HL-LHC, possibly with improved design and features, are described.

### 5.4.1 IR3 and IR7 primary collimators (target collimator primary and TCP with pick-up)

Carbon-based primary collimators, the target collimator primary (TCP), are used in the LHC to define the primary beam halo cut in the momentum (IR3) and betatron (IR7) cleaning insertion. One TCP collimator per beam is used in IR3 (horizontal orientation) whereas three collimators per beam are used in IR7 (horizontal, vertical, and skew orientations) for a total of eight primary collimators in the LHC. Since these collimators are closest to the circulating beams, their jaws are built with a robust carbon-fibre composite (CFC) that is designed to withstand the design LHC failure scenarios at injection (full injection train of 288 bunches impacting on one jaw) and at 7 TeV (up to eight bunches impacting on one jaw in the case of an asynchronous dump) [1]. The need to improve the TCP collimator design in view of the updated beam parameters for the HL-LHC design is being assessed.

The LHC primary collimator might needed to be upgraded for the HL-LHC if the present design:

- proves to not be adequate to cope with the design LHC failure scenarios updated for the upgraded HL-LHC beam parameters (larger bunch intensity and smaller emittances);
- proves to not be adequate for standard operational losses with a larger stored beam energy in the HL-LHC: for the same assumed minimum beam lifetime in operation, the total loss rates expected on the collimators might be up to a factor of 2 larger for the HL-LHC than for the LHC;
- can be improved in a way that the HL-LHC could profit from, e.g. improved materials or alignment features (integrated BPMs) for a more efficient operation.

The primary collimators are a fundamental element of the LHC multi-stage collimation hierarchy and are required in all operational conditions with beam in the machine. These are therefore high-reliability devices that must be compatible with operation in very high radiation environments and withstand standard operational losses and relevant failure cases without permanent damage that could jeopardize their functionality.

Note that a design with BPM-integrated jaw for primary collimators is currently being studied. This design, referred to as TCP with pick-up (TCPP) currently uses the same CFC materials for the jaw but provides greatly improved operational features in terms of alignment speed and beam position monitoring. This design is also being considered for HL upgrades.

### 5.4.2 IR3 and IR7 secondary collimators (target collimator secondary graphite)

Carbon-based secondary collimators (target collimator secondary graphite (TCDG)) are used in the LHC for the secondary stage of the beam halo cut in the momentum (IR3) and betatron (IR7) cleaning insertion. Four secondary collimators per beam are used in IR3 whereas 11 collimators are used in IR7 for a total of 30 TCSG collimators in the LHC. Horizontal, vertical, and skew orientations are used in different locations. Since these collimators are among the closest to the circulating beams, their jaws are built with a robust CFC that is designed to withstand the same design LHC failure scenarios at injection and at 7 TeV as the primary collimators.

The present baseline for the HL-LHC is that new secondary collimators, TCSPM, based on advanced robust and low-impedance materials will be added in IR3 and IR7, using existing TCSM slots [2]. In this scenario, the need to maintain the operability of the present CFC secondary collimators remains to be assessed.

This depends, for example, on whether the new TCSPM collimators will be able to withstand the injection failure scenario. These aspects are presently under study.

### 5.4.3 IR3 and IR7 active shower absorbers collimators (target collimator long absorber)

Tungsten-based shower absorbers collimators (target collimator long absorber (TCLA)) are used in the LHC for the third or fourth stage of cleaning of beam halos in the momentum (IR3) and betatron (IR7) cleaning insertion. Four TCLA collimators per beam are used in IR3 whereas five collimators are used in IR7, with a total of 18 TCLA collimators in the LHC. Horizontal and vertical orientations are used depending on the location. Operationally, these collimators are not supposed to intercept primary or secondary beam losses. They are therefore built using a heavy tungsten alloy that maximizes efficiency in cleaning but which is not robust with respect to a beam impact of considerable power. The need to improve the TCLA collimator design in view of the updated beam parameters for the HL-LHC design is being assessed.

As for the previous case, the upgrade of the LHC shower absorber collimators might be needed for the HL-LHC if the present design proves to not be adequate for the standard operational losses with a larger stored beam energy in the HL-LHC and/or if it can be improved in a way from which the HL-LHC could profit (improved materials, BPM features).

The TCLA collimators are an important element of the LHC multi-stage collimation hierarchy and are required in all operational conditions with beam in the machine. Operation might continue temporarily in the case of isolated TCLA failures, but we assume here that HL operation for physics without TCLA collimators will not be possible. These are therefore high-reliability devices that must be compatible with operation in very high radiation environments and withstand standard operational losses and relevant failure cases without permanent damage that could jeopardize their functionality.

A joint study by the collimation team and the beam dump team has indicated the addition of two TCLA collimators per beam in IR6 in order to improve the protection of the Q4 and Q5 magnets immediately downstream of the dump protection devices [49]. The results indicate that this improvement was not necessary for post-LS1 operation. The requirements for HL will be reviewed in 2015.

### 5.4.4 IR6 secondary collimators with pick-up (target collimator secondary with pick-up)

Carbon-based secondary collimators with pick-up buttons (target collimator secondary with pick-up (TCSP)) are used in the LHC IR6 insertion as a part of the LHC protection system. Two collimators are used in the LHC, one per beam, as auxiliary dump protection devices in the horizontal plane. In LS1, the TCSG design without integrated beam position monitors (BPMs) was replaced with the new one with BPMs for improved alignment and local orbit monitoring. Since these collimators are among the closest to the circulating beams, and are expected to be heavily exposed to beam losses in case of asynchronous dumps, their jaws are built with a robust CFC that is designed to withstand the design LHC failure scenarios at injection (full injection train of 288 bunches impacting on one jaw) and at 7 TeV (up to eight bunches impacting on one jaw in case of an asynchronous dump). The need to improve the IR6 TCSP collimator design in view of the updated beam parameters for the HL-LHC design is being assessed.

### 5.4.5 Passive absorbers in IR3 and IR7 (TCAPA, TCAPB, TCAPC, TCAPD)

Tungsten-based passive shower absorbers collimators (target collimator absorber passive (TCAP)) are used in the LHC as fixed-aperture collimators in the momentum (IR3) and betatron (IR7) cleaning insertion to reduce radiation doses to the warm quadrupole and dipoles in these insertions. Two TCAP collimators per beam are used in IR3 whereas three collimators are used in IR7 for a total of 10 TCAP collimators in the LHC. Four variants of these collimators exist to match the dimensions and orientations of the aperture of the adjacent warm magnets: TCAPA, TCAPB, TCAPC, TCAPD. Operationally, these collimators are not supposed to intercept primary or secondary beam losses but rather to absorb shower products generated by halo particles impinging on primary and secondary collimators. They are built using a heavy tungsten alloy that maximizes shower absorption, surrounded by copper. The need to improve the TCAP collimator design in view of the updated beam parameters for the HL-LHC design is being assessed.

The TCAP collimators ensure that doses on warm magnets in the cleaning insertions are minimized. Doses are determined by the integrated luminosity and therefore the possibility to improve the warm magnet protection must be envisaged for the HL-LHC luminosity goal. The upgrade of the passive absorber collimators might be needed for the HL-LHC if the present design proves not to be adequate for HL-LHC operational loss cases and/or if it can be improved by increasing the lifetime of warm magnets due to radiation wear (e.g. new materials or improved layouts/designs).

### 5.4.6 Tertiary collimators with pick-up in the experimental regions (target collimator tertiary with pick-up)

As discussed above, tungsten-based tertiary collimators with pick-up buttons (target collimator tertiary with pick-up (TCTP)) are used in the LHC to protect the superconducting triplets and the experiments in each experimental insertion against horizontal (TCTPH) and vertical (TCTPV) beam losses. A complete re-design of IR1 and IR5 collimation is imposed by the layout changes foreseen in the HL-LHC for LS3. The need to improve the present TCTP collimators in IR1/IR5 in view of the updated beam parameters for the HL-LHC design is also being assessed. In particular, consideration is being given to the possibility of replacing the TCTPs with more robust ones based on novel materials, at least in the horizontal plane, which is affected by beam dump failures.

### 5.4.7 Physics debris collimators in the experimental regions (Target Collimators Long - TCL)

Physics debris absorbers are used in IR1 and IR5 to protect the matching sections and the dispersion suppressor from beam losses caused by collision product. The LHC IR layouts as of 2015 feature three horizontal TCL collimators per beam and per IR, for a total of 12 TCL collimators, installed in cells 4, 5, and 6. Their jaws are made of copper (TCLs in cells 4 and 5) and Inermet 180 (cell 6), as the latter were installed by recuperating tertiary collimators replaced in LS1 with the new TCTPs. Indeed, the tungsten based TCT can serve as TCL without design changes. The present baseline foresees changing all TCLs for HL by adding the BPM feature. If it is proven that this new feature is not mandatory, the present TCL collimators might be re-used for the HL-LHC by being moved to new layout positions. Mechanical and radiation hardware should be studied for this scenario.